\documentclass[12pt]{article}
\newlength{\extraspace}
\newlength{\extraspaces}
\newcommand{\be}{\begin{equation}
\addtolength{\abovedisplayskip}{\extraspaces}
\addtolength{\belowdisplayskip}{\extraspaces}
\addtolength{\abovedisplayshortskip}{\extraspace}
\addtolength{\belowdisplayshortskip}{\extraspace}}
\newcommand{\ee}{\end{equation}}

\newcommand{\ba}{\begin{eqnarray}
\addtolength{\abovedisplayskip}{\extraspaces}
\addtolength{\belowdisplayskip}{\extraspaces}
\addtolength{\abovedisplayshortskip}{\extraspace}
\addtolength{\belowdisplayshortskip}{\extraspace}}
\newcommand{\ea}{\end{eqnarray}}

\newcommand{\newsection}[1]{
\vspace{12mm}
\pagebreak[3]
\addtocounter{section}{1}
\setcounter{subsection}{0}
\setcounter{footnote}{0}
\noindent{\bf \thesection. #1}
\nopagebreak
\medskip
\nopagebreak}

\newcounter{saveeqn}

\newcommand{\dif}{\mathrm{d}}
\newcommand{\me}{\mathrm{e}}

\begin{document}
\addtolength{\baselineskip}{1.5mm}

\thispagestyle{empty}
\begin{flushright}
gr-qc/0410002\\
\end{flushright}
\vbox{}
\vspace{2.5cm}

\begin{center}
{\LARGE{A new form of the rotating C-metric
        }}\\[16mm]
{Kenneth Hong~~and~~Edward Teo}
\\[6mm]
{\it Department of Physics,
National University of Singapore, 
Singapore 119260}\\[15mm]

\end{center}
\vspace{2cm}

\centerline{\bf Abstract}
\bigskip 
\noindent 
In a previous paper, we showed that the traditional form of the 
charged C-metric can be transformed, by a change of coordinates,
into one with an explicitly factorizable structure function.
This new form of the C-metric has the advantage that its properties
become much simpler to analyze. In this paper, we propose an analogous 
new form for the rotating charged C-metric, with
structure function $G(\xi)=(1-\xi^2)(1+r_+A\xi)(1+r_-A\xi)$, 
where $r_\pm$ are the usual locations of the horizons in the 
Kerr--Newman black hole. Unlike the non-rotating case, this new form 
is not related to the traditional one by a coordinate transformation. 
We show that the physical distinction between these two forms of the
rotating C-metric lies in the nature of the conical singularities 
causing the black holes to accelerate apart: the new form is free of 
torsion singularities and therefore does not contain any closed 
timelike curves. We claim that this new form should be considered
the natural generalization of the C-metric with rotation.


\newpage

\newsection{Introduction}

The charged C-metric describes a pair of Reissner--Nordstr\"om black 
holes uniformly accelerating away from each other under the action of
conical singularities in the spacetime. Its line element
has traditionally been written in the form \cite{Kinnersley:zw}:
\be
\label{CM}
\dif s^2=\frac{1}{A^2(x-y)^2}\left[G(y)\,\dif t^2-\frac{\dif y^2}
{G(y)}+\frac{\dif x^2}{G(x)}+G(x)\,\dif\phi^2\right],
\ee
with the structure function
\be
\label{old_SF}
G(\xi)=1-\xi^2-2mA\xi^3-q^2A^2\xi^4,
\ee
where $m$, $q$ and $A$ are parameters related to the mass, charge
and acceleration of the black holes, respectively. However, in a recent 
paper \cite{Hong:2003gx}, we advocated a new form of the charged
C-metric, given by (\ref{CM}) but with the explicitly factorizable 
structure function
\be
\label{new_SF}
G(\xi)=(1-\xi^2)(1+r_+A\xi)(1+r_-A\xi)\,,
\ee
where $r_\pm=m\pm\sqrt{m^2-q^2}$ are the usual locations of the
horizons of the Reissner--Nordstr\"om black hole. This new form is
related to the previous one by a simple change of coordinates, but 
it has the obvious advantage that the roots of $G(\xi)$ are now 
straightforward to write down, compared with those of (\ref{old_SF}). 
As was demonstrated in \cite{Hong:2003gx}, this leads to potentially 
drastic simplifications when analyzing the properties of the charged 
C-metric, since many results depend explicitly on these roots.

In this paper, we explore the extension of this idea to the 
rotating (charged) C-metric, which describes a pair of Kerr--Newman black
holes uniformly accelerating away from each other, again under the 
action of conical singularities in the spacetime. In its traditional form, 
the line element of this spacetime is given by \cite{Pravda:2002kj}
\ba\label{RC} &&\dif
s^2=\frac{1}{A^2(x-y)^2}\left[\frac{G(y)}{1+(aAxy)^2}\left(\dif
t-aAx^2\,\dif\phi\right)^2-\frac{1+(aAxy)^2}{G(y)}\,\dif
y^2\right.\nonumber\\
&&\qquad\qquad\qquad\qquad\left.{}+\frac{1+(aAxy)^2}{G(x)}\,\dif
x^2+\frac{G(x)}{1+(aAxy)^2}\left(\dif\phi+aAy^2\,\dif
t\right)^2\right], \ea
with the structure function
\be
\label{oldRC_SF}
G(\xi)=1-\xi^2-2mA\xi^3-(a^2+q^2)A^2\xi^4,
\ee
where $a$ is a parameter related to the the black holes' angular momentum
(per unit mass). As in (\ref{old_SF}), the roots of this quartic 
polynomial will in general be very complicated to write out explicitly. 
It is therefore worth asking if the structure function of this metric
can be written in an explicitly factorizable form. Indeed, we shall 
show that it is possible to write it in the form (\ref{new_SF}), 
but with $r_\pm$ now given by
\be
\label{def_rpm}
r_\pm=m\pm\sqrt{m^2-a^2-q^2}\,, 
\ee
which are the usual locations of the horizons of the Kerr--Newman 
black hole.

There is, however, one crucial difference with the non-rotating case.
It turns out that the new form of the rotating C-metric is physically
distinct from the one in the traditional form, and so they cannot be 
related by a 
coordinate transformation. As we shall demonstrate, the two spacetimes
differ in the nature of the conical singularities accelerating the
two black holes apart. There are in general {\it torsion singularities\/}
\cite{Letelier:1998ft,Bonnor:2001} associated with the conical 
singularities in the traditional form of
the rotating C-metric (\ref{RC}), (\ref{oldRC_SF}), while these torsion 
singularities are absent in the new form. 

Recall that torsion singularities arise when conical singularities possess
a non-zero angular velocity. It is also possible to show that there are
necessarily closed timelike curves (CTCs) in the neighbourhood of a 
torsion singularity \cite{Bonnor:2002fk}. This is pathological and thus
physically undesirable. 
Therefore, we argue that our new form of the rotating C-metric {\it is 
the correct\/} generalization of the C-metric with rotation, as it does 
not suffer from the presence of CTCs as in the traditional form (\ref{RC}), 
(\ref{oldRC_SF}).

The organization of this paper is as follows: In Sec.~2, we will review
how the traditional form of the rotating C-metric can be obtained from the 
general solution of Plebanski and Demianski \cite{Plebanski:gy}. 
At the same time, we will show how a different choice of parameters 
in this general solution leads to the new form of the rotating C-metric.
Some properties of the latter are then discussed in Sec.~3; in 
particular, we uncover the physical distinction between the traditional
and new forms of the rotating C-metric. The paper ends with a brief
discussion of how these results can be extended to include a background
electric field or cosmological constant.

\newsection{Derivation}

In 1976, a very general class of dyonic solutions to Einstein--Maxwell 
theory (including a cosmological constant) was found by Plebanski 
and Demianski \cite{Plebanski:gy}. Setting the cosmological constant
and magnetic charge to zero, the line element and electromagnetic 
potential of their solution are given by
\ba
\label{PD} 
\dif s^2&=&\frac{1}{(x-y)^2}\left[
\frac{G(y)}{1+(xy)^2}\left(\dif t-x^2\,\dif\phi\right)^2
-\frac{1+(xy)^2}{G(y)}\,\dif y^2
\right.\nonumber\\
&&\qquad\qquad~\left.{}
+\frac{1+(xy)^2}{G(x)}\,\dif
x^2+\frac{G(x)}{1+(xy)^2}\left(\dif\phi+y^2\,\dif t\right)^2\right]\,,\\
{\cal A}&=&\frac{qy\,(\dif t-x^2\,\dif\phi)}{1+(xy)^2}\,,
\label{PD_b}
\ea
where
\be
\label{PD_SF}
G(\xi)=\gamma+2n\xi-\epsilon\xi^2+2m\xi^3-(\gamma+q^2)\xi^4.
\ee
This solution contains five arbitrary parameters: $\gamma$, $\epsilon$,
$m$, $n$ and $q$. The first two parameters are non-trivially related to 
the acceleration and angular momentum of the solution, while the latter
three parameters are respectively related to its mass, `NUT' parameter 
and electric charge. These identifications were made by considering 
how some known solutions, such as the Kerr--Newman--NUT solution and 
the C-metric, can be recovered as special cases of this solution 
\cite{Plebanski:gy} (see also \cite{Kramer}).

To obtain the rotating C-metric, we rescale the coordinates and 
parameters as follows \cite{Pravda:2002kj}:
\ba
\label{limit}
&&\qquad~x\rightarrow\sqrt{aA}\,x\,,\quad
y\rightarrow\sqrt{aA}\,y\,,\quad
\phi\rightarrow\sqrt{\frac{a}{A^3}}\,\phi\,,\quad
t\rightarrow\sqrt{\frac{a}{A^3}}\,t\,,\cr
&&\gamma\rightarrow A^2\gamma\,,\quad
\epsilon\rightarrow\frac{A}{a}\,\epsilon\,,\quad
m\rightarrow-\sqrt{\frac{A^3}{a^3}}\,m\,,\quad
n\rightarrow\sqrt{\frac{A^5}{a}}\,n\,,\quad
q\rightarrow\frac{A}{a}\,q\,.
\ea
Then the solution (\ref{PD})--(\ref{PD_SF}) becomes
\ba
\label{newRC} 
\dif s^2&=&\frac{1}{A^2(x-y)^2}\left[\frac{G(y)}{1+(aAxy)^2}
\left(\dif t-aAx^2\,\dif\phi\right)^2-\frac{1+(aAxy)^2}{G(y)}\,\dif
y^2\right.\nonumber\\
&&\qquad\qquad\quad~\left.{}+\frac{1+(aAxy)^2}{G(x)}\,\dif
x^2+\frac{G(x)}{1+(aAxy)^2}\left(\dif\phi+aAy^2\,\dif
t\right)^2\right]\,,\\
{\cal A}&=&\frac{qy\,(\dif t-aAx^2\,\dif\phi)}{1+(aAxy)^2}\,,
\label{newRC_b}
\ea
where
\be
\label{newRC_SF0}
G(\xi)=\gamma+2nA\xi-\epsilon\xi^2-2mA\xi^3-(\gamma
a^2+q^2)A^2\xi^4.
\ee
By calculating the curvature invariants of this spacetime 
\cite{Letelier:1998rx}, we see that $\gamma$ and $\epsilon$ do not 
appear in them and are therefore kinematical parameters \cite{Plebanski:gy}.
On the other hand, these curvature invariants do depend on the so-called
dynamical parameters $m$, $n$, $q$, $a$ and $A$, with the latter two 
parameters determining the angular momentum and 
acceleration of the solution, respectively. If we 
choose $\gamma=\epsilon=1$ and set the `NUT' parameter $n$ to
zero, we recover the traditional form of the rotating C-metric
(\ref{RC}), (\ref{oldRC_SF}).

Since $\gamma$ and $\epsilon$ are kinematical parameters, 
we are actually free to choose their values without affecting the
physics of the solution. Hence, we shall adopt the new choice
$\gamma=1$, $\epsilon=1-(a^2+q^2)A^2$, so that the structure 
function (\ref{newRC_SF0}) becomes 
\be
\label{newRC_SF1}
G(\xi)=1+2nA\xi-\left[1-(a^2+q^2)A^2\right]\xi^2-2mA\xi^3
-(a^2+q^2)A^2\xi^4.
\ee
If we further set the value of the `NUT' parameter to be $n=m$,
we see that it is possible to cast the structure function
(\ref{newRC_SF1}) in the explicitly factorizable form:
\be
\label{newRC_SF2}
G(\xi)=(1-\xi^2)(1+r_+A\xi)(1+r_-A\xi)\,,
\ee
where $r_\pm$ are the usual locations of the horizons in the 
Kerr--Newman black hole, given by (\ref{def_rpm}). The solution 
(\ref{newRC}), (\ref{newRC_b}), with the structure function 
(\ref{newRC_SF2}), is therefore the new form of the rotating 
C-metric that we advocate; it clearly reduces to the factorized
form of the charged C-metric in \cite{Hong:2003gx} when the
rotation parameter $a$ is set to zero.

Because the traditional and new forms of the rotating C-metric
differ in the choice of the dynamical parameter $n$, they are
physically distinct spacetimes and so cannot be related by a change
of coordinates. This is in contrast to the non-rotating case
when $a=0$. In this limit, $n$ becomes a kinematical parameter,
and its choice no longer affects the physics of the spacetime.
As described in \cite{Hong:2003gx}, it is indeed possible to 
find a coordinate transformation relating the traditional and
new forms of the non-rotating charged C-metric.

We will uncover the physical meaning of the parameter $n$ in 
the next section. However, it is important to emphasize at this
stage, that $n$ does {\it not\/} have an interpretation in terms of
a NUT parameter. Recall that this identification was made in 
\cite{Plebanski:gy,Kramer} on the basis of how certain solutions 
with NUT parameter can be obtained as limiting cases of 
(\ref{PD})--(\ref{PD_SF}). It turns out that $n$ has a rather different 
physical interpretation in the general solution, before any limit 
is taken.

\newsection{Properties}

We now analyze some properties of the rotating C-metric in the new
form (\ref{newRC}), (\ref{newRC_b}) and (\ref{newRC_SF2}). Aspects of
the rotating C-metric have previously been studied in 
\cite{Farhoosh:1979tk,Farhoosh:1980a,Farhoosh:1980fc,
Letelier:1998rx,Bicak:1999sa,Pravda:2002kj}.

Let us assume, as in the non-rotating case \cite{Hong:2003gx}, that
the parameters of the solution satisfy
\be
0\le r_-A\le r_+A<1\,.
\ee
If we denote the four real roots of the structure function 
(\ref{newRC_SF2}) by 
\be
\label{roots}
\xi_1\equiv-\frac{1}{r_-A}\,,\qquad
\xi_2\equiv-\frac{1}{r_+A}\,,\qquad 
\xi_3\equiv-1\,,\qquad
\xi_4\equiv1\,, 
\ee 
then they obey $\xi_1\leq\xi_2<\xi_3<\xi_4$. In
order to have the correct spacetime signature, the $x$ and
$y$ coordinates must take the ranges $\xi_3\leq x\leq\xi_4$ and
$\xi_2\leq y\leq\xi_3$, respectively. Asymptotic infinity is
located at $x=y=\xi_3$. 
The black-hole event horizon is located at $y=\xi_2$, while the
acceleration horizon is at $y=\xi_3$. The line $x=\xi_4$ is the part
of the symmetry axis between the event and acceleration horizons,
while $x=\xi_3$ is that part of the symmetry axis joining up the
event horizon with asymptotic infinity.

As in the non-rotating C-metric, there are in general conical 
singularities along $x=\xi_3$ and $\xi_4$. If we take the angle 
$\phi$ to have period $\Delta\phi$, then the deficit angle along 
$x=\xi_i$ is
\be
\delta_i=2\pi-\alpha_i\,\Delta\phi\,, 
\ee
where
\be 
\alpha_3=1-2mA+(a^2+q^2)A^2,\qquad
\alpha_4=1+2mA+(a^2+q^2)A^2.
\ee
These expressions are considerably simpler than the corresponding 
ones in the traditional form of the rotating C-metric when 
$G(\xi)$ is given by (\ref{oldRC_SF}). However, the physics of these 
conical singularities remains unchanged.
Note that both conical singularities cannot be made to vanish at the
same time. If we choose to remove the conical deficit along
$x=\xi_4$ with the choice $\Delta\phi =2\pi/\alpha_4$, then there is
a positive deficit angle along $x=\xi_3$. This can be interpreted as
a semi-infinite cosmic string pulling on the black hole.
Alternatively, we can choose to remove the conical deficit along
$x=\xi_3$ with the choice $\Delta\phi=2\pi/\alpha_3$, resulting in a
negative deficit angle along $x=\xi_4$. This can be interpreted as a
strut pushing on the black hole. The strut continues past the
acceleration horizon, and joins up with a `mirror' black hole on the
other side of it.

The crucial difference with the non-rotating case lies in the fact
that these conical singularities will in general have torsion 
singularities associated with them. These arise when the conical
singularities possess a non-zero angular velocity, signified by
a non-vanishing $\omega\equiv g_{t\phi}/g_{tt}$ along the symmetry
axis \cite{Letelier:1998ft,Bonnor:2001}. Let us write the line 
element of the spacetime in Weyl--Papapetrou form:
\be
\label{WP}
\dif s^2=-\me^{2\lambda}(\dif t+\omega\,\dif\phi)^2 
+\me^{-2\lambda}\rho^2\,\dif\phi^2
+\me^{2(\nu-\lambda)}(\dif\rho^2+\dif z^2)\,,
\ee
and go near the symmetry axis by taking $\rho$ to be small. If $\omega$ 
is non-zero along the symmetry axis, then $\phi$ will become a timelike
coordinate sufficiently close to the axis. Since $\phi$ is periodically 
identified, this will lead to the presence of CTCs \cite{Bonnor:2002fk}.
The only situation in which these CTCs can be eliminated is when 
$\omega$ takes the same constant value along the {\it entire\/} symmetry 
axis. In this case, it is possible to perform a global transformation 
$t\rightarrow t-\omega|_{\rho=0}\,\phi$ \cite{Letelier:1998ft} to give 
a new spacetime free of torsion singularities and its associated CTCs.

Let us first calculate $\omega$ along the symmetry axis in the 
traditional form of the rotating C-metric (\ref{RC}), (\ref{oldRC_SF}). 
It can be shown that the values of $\omega$ along the two parts of 
the symmetry axis $x=\xi_3$ and $x=\xi_4$ are generally given by
\be
\omega_i=-aA\xi_i^2.
\ee
This is true for whichever form (\ref{oldRC_SF}), (\ref{newRC_SF1}) 
or (\ref{newRC_SF2}), of the structure function we use, provided
we denote $\xi_3$ and $\xi_4$ to be its two largest roots. In the case of 
(\ref{oldRC_SF}), we have $|\xi_3|\not=|\xi_4|$ and so the 
torsion singularities along both parts of the axis cannot 
be eliminated at the same time. We could choose to
eliminate the torsion singularity along $x=\xi_4$ using the
transformation $t\to t-\omega_4\phi$.
This would give a semi-infinite rotating cosmic string along $x=\xi_3$. 
Alternatively, we could eliminate the torsion singularity along 
$x=\xi_3$ using the transformation $t\to t-\omega_3\phi$. This 
would result in a rotating strut along $x=\xi_4$. In either case, 
there would be CTCs near some part of the symmetry axis.

On the other hand, we have $|\xi_3|=|\xi_4|$ for the factorizable 
structure function (\ref{newRC_SF2}) [c.f. (\ref{roots})] and hence 
the values of $\omega$ are the same along $x=\xi_3$ and
$\xi_4$. The torsion singularities along both parts of the symmetry 
axis can then be eliminated at the same time. Performing the 
transformation $t\to t+aA\phi$, the solution (\ref{newRC}), 
(\ref{newRC_b}) becomes
\ba
\label{newRC1} 
\dif s^2&=&\frac{1}{A^2(x-y)^2}\left\{\frac{G(y)}{1+(aAxy)^2}\left[\dif
t+aA(1-x^2)\,\dif\phi\right]^2-\frac{1+(aAxy)^2}{G(y)}\,\dif
y^2\right.\nonumber\\
&&\qquad\qquad\quad~\left.{}+\frac{1+(aAxy)^2}{G(x)}\,\dif
x^2+\frac{G(x)}{1+(aAxy)^2}\left[\left(1+a^2A^2y^2\right)\dif\phi+aAy^2\,\dif
t\right]^2\right\}\,,~~~~~\\
{\cal A}&=&\frac{qy\,[\dif t+aA(1-x^2)\,\dif\phi]}{1+(aAxy)^2}\,,
\label{newRC1_b}
\ea
with $G(\xi)$ still given by (\ref{newRC_SF2}). This is therefore
the main difference between the traditional and new forms of the
rotating C-metric: the latter is free of torsion singularities
(after the above transformation) and hence is not plagued 
by the presence of CTCs near the symmetry axis. For this reason,
we propose that (\ref{newRC1}), (\ref{newRC1_b}) and (\ref{newRC_SF2}) 
should be the correct and natural generalization of the C-metric 
with rotation.

It is instructive to see what happens if we consider the solution
with the structure function (\ref{newRC_SF1}) instead. This could be
regarded as a one-parameter generalization of our new form of the
rotating C-metric, and the presence of the linear term depending 
on $n$ in $G(\xi)$ will modify 
the expressions for the roots (\ref{roots}). In general,
we will have $|\xi_3|\not=|\xi_4|$ as in the traditional
form of the rotating C-metric, thus implying that this solution cannot 
be free of torsion singularities. It follows that the parameter $n$
determines the rate at which the conical singularities are rotating.
This interpretation is consistent with the observation made above
that $n$ loses its physical significance when $a=0$; indeed, in the
static limit we do not expect the conical singularities to be
rotating at all.

The zero-acceleration limit can be taken by performing the 
coordinate transformation:
\be
t=At',\qquad x=\cos\theta\,,\qquad y=-\frac{1}{Ar}\,, 
\ee 
and setting $A\rightarrow0$. In this limit, the solution 
(\ref{newRC1}), (\ref{newRC1_b}) and (\ref{newRC_SF2}) reduces to
\ba
\dif s^2&=&-\frac{r^2-2mr+q^2+a^2\cos^2\theta}{r^2+a^2\cos^2\theta}\left(\dif
t'-\frac{2mr-q^2}{r^2-2mr+q^2+a^2\cos^2\theta}\,a\sin^2\theta\,\dif\phi\right)^2\nonumber\\
&&+\frac{\left(r^2+a^2\cos^2\theta\right)\left(r^2-2mr+a^2+q^2\right)}{r^2-2mr+q^2+a^2\cos^2\theta}\,\sin^2\theta\,\dif\phi^2\nonumber\\
&&+(r^2+a^2\cos^2\theta)\,\left[\frac{\dif
r^2}{r^2-2mr+a^2+q^2}+\dif\theta^2\right]\,,\\
{\cal A}&=&-\frac{qr\,(\dif t'+a\sin^2\theta\,\dif\phi)}{r^2+a^2\cos^2\theta}\,,
\ea
which is the usual Kerr--Newman solution describing a single rotating
charged black hole. In fact, we would have obtained the same limit even 
if we had used the generalized structure function (\ref{newRC_SF1}) 
instead. This is due to the fact that the conical, as well as torsion, 
singularities disappear in the limit of zero acceleration.

We end off this section with a few other miscellaneous observations
of how the properties of the rotating C-metric will simplify when
the factorized structure function (\ref{newRC_SF2}) is used.
For example, the `angular velocity' of the black-hole event horizon 
turns out to be
\be 
\Omega_{\rm H}=\frac{1}{A}\,\frac{a}{r_+^2+a^2}\,,
\ee
which is far simpler than the corresponding expression in the
traditional form of the rotating C-metric \cite{Pravda:2002kj}. Up 
to an overall factor $1/A$, this is, in fact, identical to the angular 
velocity of the event horizon of the Kerr--Newman black hole.
Similarly, one can calculate the area of the event horizon to be
\be 
A_{\rm H}=4\pi\,\frac{r_+^2+a^2}{1-r_+^2A^2}\,.
\ee
It clearly reduces to that of the Kerr--Newman black hole when $A=0$.

Another notable advantage would be the simplification one would get when
casting the rotating C-metric in the Weyl--Papapetrou form (\ref{WP}),
following \cite{Bicak:1999sa, Pravda:2002kj}. Recall that the
rod structure of this solution consists of a rod of finite
length representing the black-hole event horizon, together with a
semi-infinite rod representing the acceleration horizon. 
As in the static counterpart \cite{Hong:2003gx}, the new form of the
rotating C-metric leads to a more elegant description of the rod
structure, in the sense that the positions of the rod ends are 
given by very simple and natural expressions. It is, in fact, 
possible to extend the results of \cite{Hong:2003gx} directly to this
case, except that we have to replace the parameter $q^2$ by $a^2+q^2$
everywhere. In particular, the finite rod representing the event 
horizon has length $2\sqrt{m^2-a^2-q^2}/A$. Up to a scale factor $1/A$, 
this is the same as the length of the corresponding rod in the
Kerr--Newman solution.

\newsection{Discussion}

In this paper, we have proposed a new form of the rotating C-metric 
(\ref{newRC1}), (\ref{newRC1_b}), with an explicitly factorizable 
structure function (\ref{newRC_SF2}), generalizing the form of the
non-rotating C-metric in \cite{Hong:2003gx}. This solution is 
physically distinct from the traditional form of the rotating C-metric, 
in that the strut and/or cosmic string causing the black holes to 
accelerate apart do not contain any torsion singularities and is hence 
free of CTCs. Because of the latter fact, we claim that this new
form should be considered the natural generalization of the C-metric 
with rotation.

Moreover, the fact that the structure function is factorizable 
means that its roots are now straightforward to write down, and this 
leads to simplifications in the analysis of the rotating C-metric. 
For example, simple expressions are found for the angular velocity 
and area of the black-hole event horizon. This new form of the 
solution also gives a natural description of the rod structure when 
it is cast in Weyl--Papapetrou form.

There are a few possible generalizations of this new form of
the rotating C-metric. One is to consider embedding this solution 
in a background electric field, which would give a rotating 
version of the Ernst solution \cite{Ernst:1976b}. Recall that
the latter can be made free of conical singularities by 
choosing an appropriate strength of the background electric
field. In this case, it describes a pair of Reissner--Nordstr\"om
black holes accelerating apart purely under the influence of the 
background field. A natural question at this stage would be whether 
a background electric field can be introduced into the solution 
(\ref{newRC1}), (\ref{newRC1_b}) and (\ref{newRC_SF2}), to give 
a spacetime free of conical singularities. Such a solution could
be used to calculate the pair-production rate of Kerr--Newman
black holes by a background electric field, following 
\cite{Gibbons,Garfinkle:1990eq,Dowker:1993bt,Booth:1998pb}.

We have managed to obtain this generalized solution following the 
procedure described in \cite{Ernst:1976a,Ernst:1976c}. Unfortunately, 
it is given by exceedingly complicated expressions and so will not be
reproduced here. It is possible to show that this solution continues 
to be free of torsion singularities along the symmetry axes. For 
specific values of the parameters of the solution, we have checked that
it is indeed possible to remove the remaining conical singularities 
with an appropriate choice of the background electric field strength.

Recall that the original solution of Plebanski and Demianski 
\cite{Plebanski:gy} also contains a cosmological constant $\Lambda$.
It is therefore easy to embed the rotating C-metric solution in a 
de Sitter or anti-de Sitter background. A form of this solution which 
extends the factorizable structure of (\ref{newRC1}), (\ref{newRC1_b}) 
and (\ref{newRC_SF2}) is
\ba
\dif
\label{newRC_Lambda}
s^2&=&\frac{1}{A^2(x-y)^2}\left\{\frac{F(y)}{1+(aAxy)^2}\left[\dif
t+aA(1-x^2)\,\dif\phi\right]^2-\frac{1+(aAxy)^2}{F(y)}\,\dif
y^2\right.\nonumber\\
&&\qquad\qquad\quad~\left.{}+\frac{1+(aAxy)^2}{G(x)}\,\dif
x^2+\frac{G(x)}{1+(aAxy)^2}\left[\left(1+a^2A^2y^2\right)\dif\phi+aAy^2\,\dif
t\right]^2\right\}\,,~~~~~\\
{\cal A}&=&\frac{qy\,[\dif t+aA(1-x^2)\,\dif\phi]}{1+(aAxy)^2}\,,
\ea
where the structure functions are
\ba 
&&G(x)=(1-x^2)(1+r_+Ax)(1+r_-Ax)\,,\\
&&F(y)=G(y)+\frac{\Lambda}{3A^2}(1+a^2A^2y^4)\,.
\label{H}
\ea
Here, the parameters $r_\pm$ are defined by
\be
r_\pm=m\pm\sqrt{m^2-q^2-a^2\left(1+\frac{\Lambda}{3A^2}\right)}\,,
\ee
although it is important to note that they are no longer the locations 
of the horizons of the Kerr--Newman--dS/AdS black hole, as in the case 
of zero cosmological constant. In the non-rotating limit, this solution 
is equivalent to the dS/AdS C-metric solutions in \cite{Mann:1995vb,
Podolsky,Podolsky:2002nk,Dias:2002mi,Dias:2003xp}, up to a coordinate 
transformation of the form described in \cite{Hong:2003gx}.

It can be checked that the general solution (\ref{newRC_Lambda})--(\ref{H})
is free of torsion singularities along the symmetry axes $x=\pm1$, and 
should therefore be considered the natural generalization of the 
rotating C-metric with a cosmological constant. It would be interesting 
to compare the properties of this solution to the rotating dS C-metric 
solution presented in \cite{Booth:1998gf}, and also study the pair 
production of rotating black holes in the dS/AdS background
\cite{Mann:1995vb,Booth:1998gf,Dias:2003st,Dias:2004rz}.

\bigskip\bigskip

{\renewcommand{\Large}{\normalsize}
}
\end{document}